\documentclass[10pt,conference]{IEEEtran}
\IEEEoverridecommandlockouts
\usepackage{epsfig}
\usepackage{amsmath,amssymb,amstext,amsthm}
\usepackage{epstopdf}
\usepackage{graphicx} 
\usepackage{ifthen}
\usepackage{psfrag}
\usepackage{cite}
\usepackage{blindtext}
\usepackage{algorithm,algorithmic}
\usepackage{hyperref}
\usepackage{xcolor}
\usepackage{caption}
\usepackage{xspace} 
\usepackage{algorithmic}
\usepackage{tabularx,ragged2e,booktabs,caption}
\makeatletter
\g@addto@macro\normalsize{%
 \setlength\abovedisplayskip{2pt}
 \setlength\belowdisplayskip{2pt}
 \setlength\abovedisplayshortskip{2pt}
 \setlength\belowdisplayshortskip{2pt}
}
\allowdisplaybreaks

\usepackage{balance}
\usepackage[mathscr]{euscript}
 \let\mathscr\relax
\usepackage[scr]{rsfso}

\allowdisplaybreaks
\usepackage{bbm}

\newcommand{\sfr}{\mathsf{r}}
\newcommand{\st}{\mathsf{s.\ t.}}
\newcommand{\sfd}{\mathsf{d}}
\newcommand{\bfH}{\mathbf{H}}
\newcommand{\bfh}{\mathbf{h}}

\newcommand{\bfX}{\mathbf{X}}
\newcommand{\bfx}{\mathbf{x}}

\newcommand{\bfd}{\mathbf{d}}

\newcommand{\bftheta}{\boldsymbol{\theta}}
\newcommand{\bfTheta}{\boldsymbol{\Theta}}

\PassOptionsToPackage{bookmarks={false}}{hyperref}
\begin{document}
\title{\huge Performance of Joint Symbol Level Precoding and RIS Phase Shift Design in the Finite Block Length Regime with Constellation Rotation}
\author{Progress Zivuku, Steven Kisseleff, Wallace A. Martins, Hayder Al-hraishawi, \\Symeon Chatzinotas, and Bj{\"o}rn Ottersten\\
\textit{Interdisciplinary Centre for Security, Reliability and Trust (SnT), University of Luxembourg}\\ E-mails: \{progress.zivuku, steven.kisseleff, wallace.alvesmartins, hayder.al-hraishawi\\ symeon.chatzinotas, bjorn.ottersten\}@uni.lu
\thanks{This work is funded by the Luxembourg National Research Fund (FNR) as part of the CORE programme under project RISOTTI C20/IS/14773976.}
}
\maketitle

\begin{abstract}
In this paper, we tackle the problem of joint symbol level precoding (SLP) and reconfigurable intelligent surface (RIS) phase shift design with constellation rotation in the finite block length regime. We aim to increase energy efficiency by minimizing the total transmit power while satisfying the quality of service constraints. The total power consumption can be significantly minimized through the exploitation of multiuser interference by symbol level precoding and by the intelligent manipulation of the propagation environment using reconfigurable intelligent surfaces. In addition, the constellation rotation per user contributes to energy efficiency by aligning the symbol phases of the users, thus improving the utilization of constructive interference.  The formulated power minimization problem is non-convex and correspondingly difficult to solve directly. Hence, we employ an alternating optimization algorithm to tackle the joint optimization of SLP and RIS phase shift design. The optimal phase of each user's constellation rotation is obtained via an exhaustive search algorithm. Through Monte-Carlo simulation results, we demonstrate that the proposed solution yields substantial power minimization as compared to  conventional SLP, zero forcing precoding
with RIS as well as the benchmark schemes without RIS. 
\end{abstract}
\begin{IEEEkeywords}
Symbol level precoding, reconfigurable intelligent surface, energy efficiency, finite block length, constellation rotation, short packets. 
\end{IEEEkeywords} 
\section{Introduction} \label{sec:Introduction}
Current and  future generations of wireless networks are challenged to cope with the ever-increasing demands spurred by our increasingly connected world.
Such demands include high energy efficiency and data rates for a large number of users demanding access to the network\cite{yang2021energy}. 
As a consequence, there is a clear need to further advance research for new communication technologies so as to ensure that sufficient quality of service (QoS) requirements are met for various applications.

One of the main promising technologies for future wireless networks is reconfigurable intelligent surfaces (RIS). RIS is a  software-controllable meta-surface that consists of a number of passive reconfigurable reflecting elements\cite{renzo2019smart}. RIS enhances wireless communications by intelligently manipulating
the wireless propagation environment in a cost-effective and hardware
efficient manner. 

Most existing works on RIS studied the minimization of total consumed power \cite{wu2019beamforming}, network sum rate maximization \cite{huang2018achievable}, maximization of the number of served users \cite{zivuku2022maximizing}, enhancement of physical layer security \cite{makin2022enhancing} and RIS-enabled interference nulling \cite{jiang2022interference}.

In practice, multiuser interference (MUI) remains one of the limiting factors in the design of wireless networks. The capacity and QoS offered to the end users are both significantly influenced by interference. In this regard, a number of precoding techniques were proposed over the years to reduce the effects of MUI on the overall network performance \cite{joham2005linear}. Recently, symbol level precoding (SLP) gained attention due to its ability to convert the energy from interfering signals into useful signal energy by aligning the phases of the received signals in a constructive manner. This can result in improved energy efficiency and better QoS \cite{alodeh2015constructive,alodeh2018symbol,li2020tutorial}.  Recent research carried out in \cite{kisseleff2021symbol, alodeh2020joint, 9984648} has revealed that rotating each user's symbol constellation can be regarded as a new degree of freedom for SLP-based multiuser systems. Specifically, the probability of attaining constructive interference can be increased by optimizing the phase offset of each user's constellation. As pointed out in \cite{kisseleff2021symbol}, the enhancement of the energy efficiency can be especially large in combination with short packet transmission. Further, research carried out in \cite{liu2020joint} indicates the advantages of symbol level precoding in RIS-assisted networks. By exploiting MUI and the propagation environment, total transmit power can be minimized. However, this work did not take into consideration the impact of a finite packet length, which is the basis of a number of practical communication systems. Moreover, the transmission of data is packeted and the majority of practical communication systems depend on finite block length.   

Motivated by the preceding discussion, we carry out a study on joint optimization of SLP and RIS phase shift design with constellation rotation in the finite block length regime. This study aims to show the benefits of jointly exploiting both MUI and modification of the propagation environment taking into consideration the general packet length. Further, we aim to improve the probability of attaining constructive interference by obtaining the optimal rotation of each user's constellation for RIS-assisted SLP.
The major contributions of this paper are outlined as follows:
\begin{figure}[t]
	\centering	\includegraphics[width=0.45\textwidth]{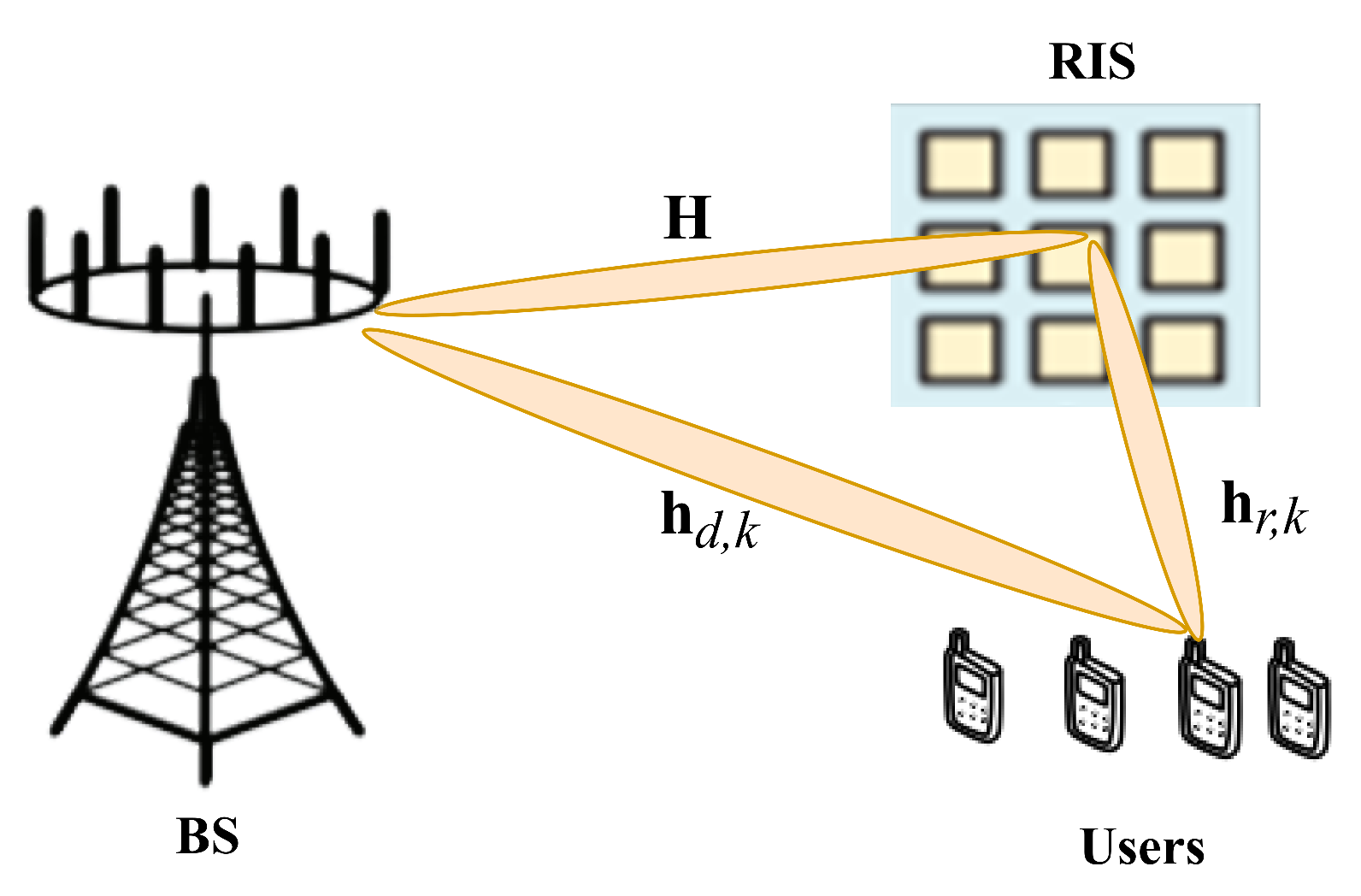} 
	\caption{RIS-assisted multi-user MISO system.}
	\label{fig:System}
\end{figure}
\begin{itemize}
    \item We investigate the benefits of SLP precoding, RIS phase shift optimization and constellation rotation mainly focusing on power minimization. We fill the gap in the current studies of joint optimization of RIS phase shifts and SLP, where a number of authors focused on the design in the infinite block length regime without constellation rotation \cite{shao2020minimum,liu2021intelligent,cheng2021degree}. In this case, we optimize the users' constellation rotation for better alignment of the symbols. As a result, the symbols can be pushed deeper into the detection region leading to higher energy efficiency. 
    \item We formulate a novel joint optimization of SLP, RIS phase shift design with constellation rotation for power minimization in the finite block length regime. The resulting problem is non-convex making it quite challenging to find a global optimal solution. In addition, there is a strong coupling between the SLP and RIS phase shift variables. Further, the discrete nature of the set of possible constellation rotations poses a significant challenge in finding an optimal solution to the problem. To tackle the complexity posed by the formulated problem, we leverage the alternating optimization technique to decouple the SLP design and the design of RIS phase shifts into two subproblems. Further, we obtain the optimal phase of constellation rotation via an exhaustive search algorithm.
    \item We provide a performance evaluation of the proposed design for a smart city street scenario under various conditions. The simulation results validated the performance improvement of the proposed design over the conventional SLP, zero forcing precoding with RIS as well as the benchmark schemes without RIS.
\end{itemize}
The remainder of the paper is organized as follows. Section~\ref{sec:SM} describes briefly the system model. In Section~\ref{sec:PM}, we present the optimization problem formulation. In Section~\ref{sec:Proposal}, we propose an approach to solve the formulated optimization problem. Section~\ref{sec:Results} presents the numerical results to evaluate the performance of the proposed design. Finally, we conclude the paper in Section~\ref{sec:Conclusion}.

Notation: Matrices and column vectors are represented by bold upper and lower case letters, respectively. $\Re\{x\}$ and $\Im\{x\}$ respectively represents the real part and the imaginary part of $x$. $\|.\|$ denotes the Euclidean norm.
\section{System Model}\label{sec:SM}

As illustrated in Fig.~\ref{fig:System}, we consider a RIS-assisted multiuser multiple-input single-output (MISO) system in the downlink, where a base station (BS) equipped with $M$ antennas serves $K$ users equipped with a single-antenna. The RIS is equipped with $N$ reflective elements installed on the wall of a high-rise building.
There are three different channels in RIS-assisted networks, which include the direct link from BS to user $k$, the indirect link from BS to RIS, and the reflection channel from RIS to user $k$ denoted by $\bfh_{\sfd,k} \in \mathbb{C}^{1\times M}$, $\mathbf{H} \in \mathbb{C}^{N\times M}$,  and $\bfh_{\sfr,k} \in \mathbb{C}^{1\times N}$, respectively.
The transmitted symbols $\bfd[t]=[d_1[t],...,d_K[t]]$ are drawn from a $\Psi$-phase shift keying (PSK) constellation, where $t$ is the index of the symbol interval. Note that we assume a quadrature PSK modulation, \textit{i.e.}, $\Psi=4$, in this work. However, the generalization to higher-order modulations is straightforward and does not impact the proposed system design.
 The BS changes its transmitted precoder $\bfx[t] \in \mathbb{C}^{M\times 1}$ with respect to different $\bfd[t]$ such that MUI can be effectively exploited. Additionally, we consider transmission over  quasi-static block-fading channels. The received signal $   r_k[t]$, at user $k$ can be written as 
\begin{equation}
    r_k[t]=\bfh_k\bfx[t]+n_k[t],
\end{equation}
where $n_k[t]$ is the additive white Gaussian noise (AWGN) with distribution $\mathcal{CN}(0,\sigma_k^{2})$. Further, $\bfh_k$ denotes the equivalent link comprised of the direct BS link and the cascaded RIS link, being expressed as
\begin{equation}
    \bfh_k=(\bfh_{d,k}+\bfh_{r,k}\bfTheta\bfH),
\end{equation} 
where $\bfTheta=\beta{\rm diag}([e^{j\theta_1} \cdots e^{j\theta_n} \cdots e^{j\theta_N}])$ is the passive reflection beamforming matrix at the RIS. In this case, $\beta \in [0 
, 1]$ is the reflection efficiency, which is equal to $1$ for simplicity. Next, $ \theta_n\in[0,2\pi] $ is the phase of the $n{\rm th}$ reflective element of the RIS.

To facilitate transmission with a finite block length $L$, the transmitted precoder matrix can be written as follows, $\bfX=[\bfx[1] \cdots \bfx[L]]$, where $\bfX$ consists of all the data vectors in the block. In addition, it is important to note that the precoder $\bfx[\ell],  \ell \in L$ changes with respect to the transmitted symbols while the phases of the RIS remain fixed within a specific coherence time, for which we assume unaltered channel state information (CSI).  \footnote{In this paper, we assume the availability of perfect CSI in order to concentrate on the impact of joint SLP, RIS phase shift design with constellation rotation in the finite block length regime. Future studies will address the impact of 
imperfect CSI on the system performance.} 

\section{Optimization Problem Formulation}\label{sec:PM}
The main objective of this work is to minimize the total transmit power by jointly optimizing SLP, constellation rotation and RIS phase shift design in the finite block length regime. Accordingly, the power minimization problem can be formulated as follows:
\begin{subequations} \label{OP1}
	\begin{IEEEeqnarray}{cl}
\mathscr{P:}\quad	   &\underset{\bfX,\bftheta,\boldsymbol{\phi} }{\mathrm{minimize}} \   \|\bfX\|^2\label{OP1_a} \\
		& \st  \quad\ \Re{(\bfh_k{\bfx}[\ell])}-\sigma_k\sqrt{\gamma_k}\Re{(e^{-j\phi_k}\cdot d_k[\ell])}\unlhd 0, \nonumber \\ &\qquad\quad \forall k,\ell \ \label{OP1_b} \\
        & \qquad\quad\ \Im{(\bfh_k{\bfx}[\ell])}-\sigma_k\sqrt{\gamma_k}\Im{(e^{-j\phi_k}\cdot d_k[\ell])}\unlhd 0, \nonumber \\ &\qquad\quad \forall k,\ell \ \label{OP1_c} \\
		     & \qquad\quad  \phi_k \in \left\{0,\frac{2\pi}{\Psi_k},...,(\Psi_k-1)\frac{2\pi}{\Psi_k}\right\}, \forall k, \label{OP1_d}\ \\
		     & \qquad\quad  0\le\theta_{n}\le2\pi, \forall n  \  \label{OP1_e}\	\end{IEEEeqnarray}
\end{subequations}
where, $\unlhd$ is an element-wise operator that guarantees receiving each symbol in the correct detection region.
Further details on the extended regions are well presented in \cite{li2020tutorial}.
In addition, it is important to note that the received symbol constellation is scaled by $\sigma_k\sqrt{\gamma_k}$, where $\sigma_k$ denotes the noise standard-deviation and $\gamma_k$ is the desired user's signal-to-interference-plus-noise ratio (SINR).
The constraints considered in this problem are explained in detail as follows:
\begin{itemize}
    \item Constraints~\ref{OP1_b} and ~\ref{OP1_c}: These are bi-linear constraints that ensure the received signals fall in the correct detection regions. Note that these regions are defined according to the selected quadrature PSK modulation.
    \item Constraint~\ref{OP1_d}: This constraint describes the discrete set of possible phases of the users' constellation rotation.
    \item Constraint~\ref{OP1_e}: This constraint describes the RIS phase shifts.
\end{itemize} 
It is important to note that problem $\mathscr{P}$ in~(3) is non-convex. Further, there is a strong coupling between discrete and continuous variables. In this case, the problem cannot be solved using standard convex optimization techniques, making it computationally intractable. To tackle the complexity of the formulated optimization problem, we split the joint optimization of SLP and RIS phase shift design into two tractable sub-problems. The two sub-problems can be solved in an alternating manner by leveraging the alternating optimization framework. Further, the optimal phase of the constellation rotation for each user is obtained via an exhaustive search algorithm. This is motivated by the fact that a slight change in the constellation rotation of one of the users leads to a completely different operation point in the parameter space since this rotation affects the shape of the region in a non-linear way and the phase offset of each rotation is drawn from a rather small set of values as suggested in \cite{kisseleff2021symbol}. Accordingly, such changes make the optimization procedure unstable, \textit{i.e.} no stationary solution can be guaranteed while the performance is typically poor and the complexity of optimization is very high. In order to avoid this pitfall, we perform the optimization of the precoder and the phase shifts for each combination and then select the best combination via exhaustive search. \footnote{The proposed design can be enhanced by applying low-complexity methods proposed in \cite{9984648} instead of exhaustive search. However, the integration of this method in our proposed joint optimization of SLP, RIS phase shifts and constellation rotation is beyond the scope of this work.}

\section{Proposed Optimization Algorithm}\label{sec:Proposal}
In this section, we detail the steps of the optimization procedure outlined above.

Specifically, for each combination of rotation, \footnote{The number of symbol combinations $Q$ is equal to $\Psi^K$. Assuming  quadrature PSK modulation, the total number of combinations of constellation rotations is also $Q$ as suggested in \cite{kisseleff2021symbol}.} we optimize the SLP and the RIS phase shift design for all of the block symbols via an alternating optimization framework. In this case, we obtain a combination with the minimum consumed total power. It is important to note that the BS changes the precoder $\bfx[\ell]$ for all the block symbols while the RIS phase shifts remain fixed. Specifically, the RIS phase shifts are optimized considering all the different symbol combinations in a block of length $L$.

\subsection{Symbol level precoding optimization}
In this subsection we tackle the design of transmit precoder vectors  $\bfx[1],\cdots,\bfx[L]$. Specifically, for a given RIS phase shift vector $\bftheta$ and given combination of constellation rotations, we solve the SLP problem to obtain the transmitted signal $\boldsymbol{x}[\ell]$ for all symbol combinations in a block. The resultant subproblem is formulated as follows,
\begin{subequations} \label{OP2}
	\begin{IEEEeqnarray}{cl}
\quad	   &\underset{\bfx[\ell], \forall \ell}{\mathrm{minimize}} \   \|\bfx[\ell]\|^2\label{OP2_a} \\
	& \st  \quad\ \Re{(\bfh_k{\bfx}[\ell])}-\sigma_k\sqrt{\gamma_k}\Re{(d_k[\ell])}\unlhd 0,  \forall k \ \label{OP2_b} \\
            & \qquad \quad\ \Im{(\bfh_k{\bfx}[\ell])}-\sigma_k\sqrt{\gamma_k}\Im{(d_k[\ell])}\unlhd 0, \forall k \ \label{OP2_c} 
\end{IEEEeqnarray}
\end{subequations}
The problem in (4) is convex. In this case, standard convex tools, e.g., CVX solver \cite{ben2001lectures} can be exploited to solve the problem effectively. In addition, the problem can be solved using a non-negative least squares (NNLS) algorithm, which uses the geometry of SLP constructive regions\cite{haqiqatnejad2019approximate}. For more details of the NNLS algorithm, refer to \cite{bro1997fast}.

\subsection{RIS Phase Shift Design}
In this subsection, we tackle the design of RIS phase shifts for given transmit precoder vectors. It is important to note that after solving the SLP problem in (4), the optimization of RIS phase shifts will not result in further reduction of total transmit power. Accordingly, with the phase shift optimization, we aim at increasing the parameter subspace by making the QoS constraints, i.e. received signal constraints, inactive, such that the transmit power can be subsequently reduced via precoder design in the next iteration. For that reason, to further minimize the transmit power in the next iterations, we model a new objective function to maximize the QoS. In this case, the phases of the RIS are optimized to improve the QoS which can then increase the parameter space around the previously found solution. The resultant optimization problem is formulated as follows,
\begin{subequations} \label{OP3}
	\begin{IEEEeqnarray}{cl}
\quad	   &\underset{\bftheta, \mathbf{Z}}{\mathrm{maximize}} \   \sum_{\ell=1}^L\sum_{k=1}^k(z_{\ell,k}) \label{OP3_a} \\
		& \st  \quad\ \Re{(\bfh_k{\bfx}[\ell])}-z_{\ell,k}\Re{(d_k[\ell])}\unlhd 0, \forall k,\ell \ \label{OP3_b} \\
            & \qquad \quad\ \Im{(\bfh_k{\bfx}[\ell])}-z_{\ell,k}\Im{(d_k[\ell])}\unlhd 0,  \forall k,\ell \ \label{OP3_c} \\
		     & \qquad\quad  0\le\theta_{n}\le2\pi, \forall n  \  \label{OP3_d},\	\end{IEEEeqnarray}
\end{subequations}
where, $(\ell,k)$-th entry of $\mathbf{Z}$ is $z_{\ell,k}$, which is the QoS metric. In this case, $z_{\ell,k}=\sigma_{k}\sqrt{\gamma_k}$. A similar procedure has been proposed in \cite{liu2020joint}.
We note that the problem formulated in (5) is convex and can be solved via convex optimization tools, e.g CVX solver \cite{ben2001lectures}. Considering the two formulations in subproblems (4) and (5), the joint symbol level precoding and RIS phase shift design can be easily solved via alternating optimization. In this case, the precoder $\bfX$ and RIS phase shifts $\bftheta$ are iteratively updated until convergence. We note that solving the joint optimization of SLP and RIS phase shift design requires an initial feasible point $\bftheta^{(0)}$. Accordingly, we  generate $\bftheta^{(0)}$ as an identity phase shift matrix.

The joint SLP and phase shift design is carried out for different combinations of constellation rotations and we obtain the combination with the minimum total transmit power. We summarize the joint  SLP, RIS phase shift and constellation rotation optimization in Algorithm~\ref{alg1}.
\begin{algorithm}[h!]
\begin{algorithmic}[1]
 \fontsize{9.5}{9.5}
\selectfont
\protect\caption{: Joint SLP and RIS phase shift design in the finite block length with constellation rotation}
\label{alg1}
\global\long\def\algorithmicrequire{\textbf{Input:}}
\REQUIRE  $\bfH, \bfh_{r,k},\bfh_{d,k},L, \sigma_k, \gamma_k,\bftheta^{(0)}$
\global\long\def\algorithmicrequire{\textbf{Output:}}
\REQUIRE $\bftheta^{\star},\bfX^{\star}$
\STATE Set $P_{\rm rotation}[m]=0, \forall m \in Q$ 
\STATE \textbf{for} $m=1$ \textbf{to} $Q$ \textbf{do}
\STATE Get the phase $\phi_k$, $\forall k$ for the $m^{\rm th}$ combinations of rotations.
\STATE Set $\varrho=0$
\REPEAT
\STATE \textbf{for} $\ell$=1 \textbf{to} $L$
\textbf{do}
\STATE solve \eqref{OP2} for a given $\bftheta^{\varrho}$ to obtain optimal solution ($\bfx^{\star}[\ell]$) 
\STATE \textbf{end for}
\STATE update $\bfX^{(\varrho+1)} \leftarrow \bfx[\ell],\cdots, \bfx[L]$
\STATE Solve  \eqref{OP3} for given $(\bfX^{(\varrho+1)})$ to obtain the optimal solution $(\bftheta^{\star})$   \STATE update $(\bftheta^{(\varrho+1)}):=  (\bftheta^{\star})$;
\STATE Set $\varrho=\varrho+1$;
\UNTIL Convergence
\STATE $ P_{\rm rotation}[m] \leftarrow\|\bfX\|^2$;
\STATE \textbf{end for}
\STATE $\bfX$ =$ {\rm argmin}P_{\rm rotation}[m]$
\STATE $\bftheta^{\star}=\bftheta$, $\bfX^{\star}=\bfX$.
\end{algorithmic} \end{algorithm}

\subsection{Complexity Analysis} We briefly provide a complexity analysis of the proposed optimization algorithm. According to \cite{bro1997fast}, the complexity of problem in (4) by the NNLS algorithm is given by $\mathcal{O}\bigl(MK^2)$. Next, the complexity of problem in (5) is given by $\mathcal{O}\bigl(\sqrt{(N+2KL)}(N+KL)^3)$. The overall complexity of the proposed design is $\mathcal{O}\bigl(Q\boldsymbol{\varrho}(MK^2+\sqrt{(N+2KL)}(N+KL)^3))$. As previously mentioned, the complexity of the proposed design can be reduced by implementing low-complexity algorithm proposed in \cite{9984648}. 
\section{Numerical Results and Discussion}\label{sec:Results}
In this section, we conduct Monte-Carlo simulations to evaluate the performance of the proposed solution against the state of the art benchmark schemes. We consider a 3-D scenario where a BS is deployed at the origin and the RIS is deployed $x$ distance from the base station on the bulding facade of a city street. The users are uniformly distributed at random between $20$ to $40$ meters in the  y-direction, see Fig.~\ref{fig:System}. Due to the location of the RIS which is high above the users, the pathloss exponent $\rho_{\textrm{br}}$ of the BS-RIS link is set lower than that of the other links. The symbol constellation scaling is set at $\sigma_k\sqrt{\gamma_k}= 10^{-7}$. This can be adjusted according to different QoS requirements.

We modeled the distance-dependent
path loss as $\alpha=-30-10\rho\log_{10}(d)$~dB \cite{9110587}, where $d$ 
is the link distance, and $\rho$ is the path-loss exponent. To account for small scale fading, we considered a Rician channel model for all channels involved. For instance, the BS-user $k$ channel $\bfh_{d,k}$ can be written as $\bfh_{d,k}=\sqrt{\frac{\eta}{1+\eta}}\bfh_{d,k}^{\rm LOS}+\sqrt{\frac{1}{1+\eta}}\bfh_{d,k}^{\rm NLOS}$, where $\bfh_{d,k}^{\rm LOS}$ and $\bfh_{d,k}^{\rm NLOS}$ denote the LOS and NLOS component, respectively. The Rician factor $\eta$ is $10$~dB. To validate the performance of the  proposed joint SLP and RIS phase shift design with constellation in the finite block length regime (Proposed), we compared it with $5$ benchmark schemes listed below:
\begin{itemize}
    \item Zero forcing precoding without RIS (ZF, no RIS): In this scheme ZF precoding is performed without the help of the RIS.
    \item Zero forcing precoding with RIS (ZF, with RIS): In this scheme ZF precoding is performed where transmission between the users and the BS is aided by the RIS.
    \item Conventional symbol level precoding without RIS (SLP infinite, no RIS): In this scheme we perform conventional SLP precoding without the help of the RIS.
    \item Conventional symbol level precoding with RIS (SLP infinite, with RIS): In this scheme we perform conventional SLP precoding with the help of the RIS.
    \item Proposed solution without RIS. (SLP finite, no RIS): In this scheme we carry out SLP precoding in the finite block length with constellation rotation. However, communication is carried out via the BS direct link.
\end{itemize}
For all simulation results in this paper, we average the transmit power in 100 channel realizations and 10 blocks per realization which results in 1000 blocks. All the analyses in this paper were carried out focusing on the QPSK modulation. Parameters utilized in the simulations are specified in Table~\ref{tab:Simulationparameter}.
\begin{table}[h!]
	\centering  
	\captionof{table}{Simulation Parameters}	\label{tab:Simulationparameter}
	 \scalebox{1.1}{
		\begin{tabular}{l|l}
			\hline
			Parameter & Value \\
			\hline\hline
			 Carrier frequency $\left(f_{c}\right)$ & $2.4 ~\mathrm{GHz}$ \\
              Number of users ($K$) & 4 \\
              Number of BS antennas ($M$) & 4 \\ 
              Number of RIS elements ($N$) & 64\\
			 BS height  & $3~ \mathrm{~m}$ \\
			 User height  & $1.5 \mathrm{~m}$ \\
			 RIS height  & $3~ \mathrm{~m}$ \\
			 RIS $x$ distance  & $3~ \mathrm{~m}$ \\
			 pathloss exponent BS-RIS ($\rho_{\textrm{br}}$) & 2.3 \\
              pathloss exponent RIS-user ($\rho_{\textrm{ru}}$) & 2.6 \\
              pathloss exponent BS-user ($\rho_{\textrm{bu}}$) & 2.6 \\ 
             Antenna and element spacing & $0.5\lambda$ \\
			\hline		  		
		\end{tabular}
	}
\end{table}
\subsection{Analysis of average transmit power as a function of block length}
\begin{figure}[h!]
    \centering	\includegraphics[width=0.5\textwidth]{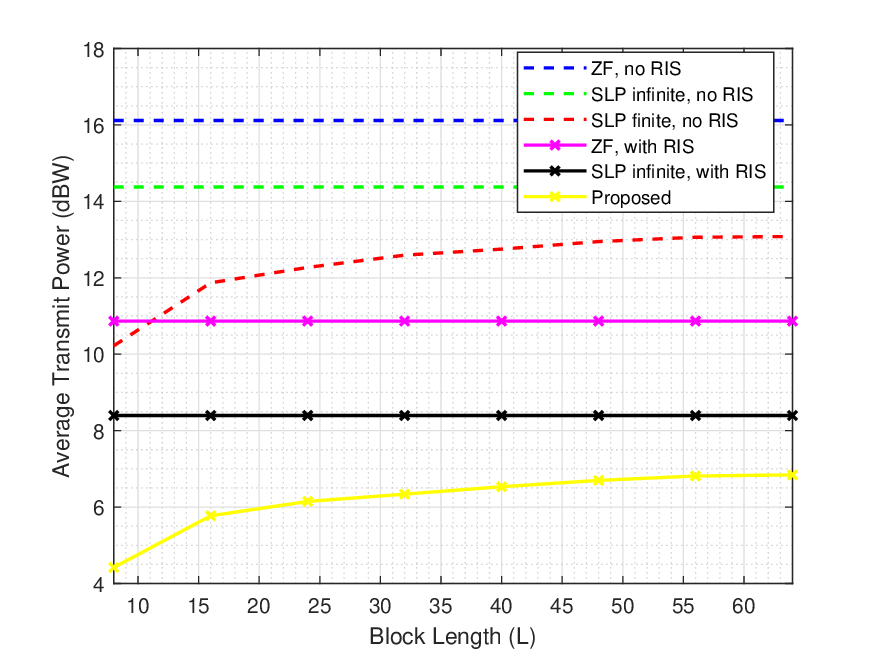}
	\caption{Average transmit power as a function of block length.}
	\label{fig:Const}
\end{figure}
In this subsection, we study the effect of block length on the average transmit power. We vary the block length $L$ from $8$ to $64$ and the results are depicted in Fig.~\ref{fig:Const}.  The proposed solution with RIS and constellation rotation results in a substantial reduction in power consumption as compared to all the considered benchmark schemes. We observe higher gains of the proposed design against conventional SLP precoding (SLP infinite, with RIS) when the block length is short.  Specifically, at $L=8$, the gain between the proposed solution and the conventional SLP is about $4$~dB. In this case, the design of RIS-assisted SLP with constellation rotation is important for reduced power consumption in the finite block length regime. Further, the proposed design results in power savings of about $5.5$~dB with respect to the design without RIS (SLP finite, no RIS). This validates the impact of our proposed scheme in reducing the total consumed power as compared to the design without RIS. This also shows the symbiotic benefits of MUI exploitation by SLP and the manipulation of the propagation environment by RIS. It is important to note that with increasing block length, the transmit power for the SLP in finite block length regime increases. However we still note a performance gain of about $1.5$~dB with respect to the conventional SLP at $L=64$. Further, we obtain a performance gain of $6$~dB at $L=64$ with the proposed solution compared to the finite block length SLP with no RIS.  
\subsection{Analysis of average transmit power as a function of the number of RIS elements}
\begin{figure}[h!]
	\centering	\includegraphics[width=0.5\textwidth]{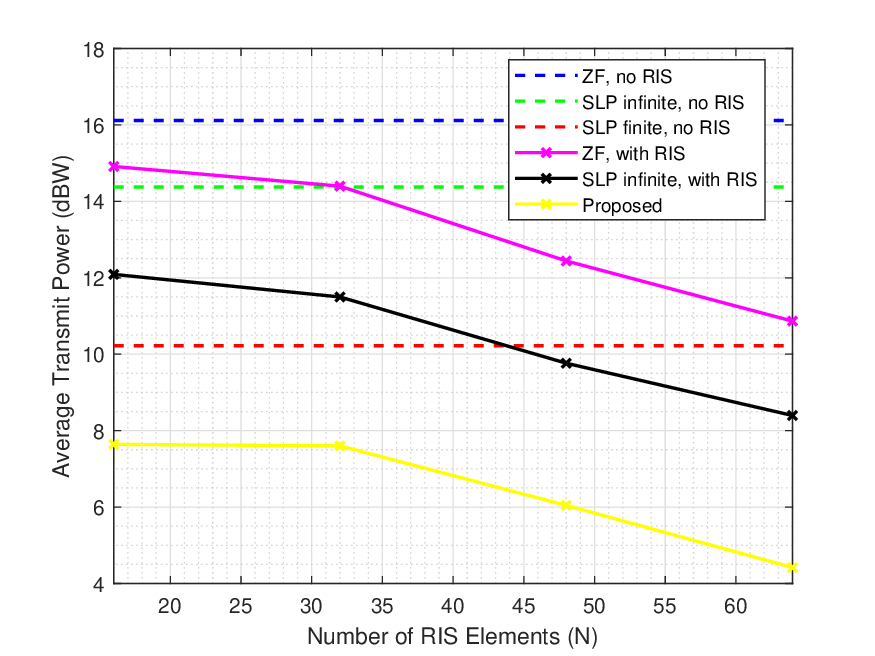}
	\caption{Average transmit power as a function of the number of RIS elements.}
	\label{fig:Con}
\end{figure}
In this subsection, we analyze the effect of the number of RIS elements on the average total transmit power. For this analysis, we set the block length to $L=8$ and the number of RIS elements is varied between $16$ and $64$. The proposed solution provides significant power savings compared to all the considered benchmark schemes as shown in Fig.~\ref{fig:Con}. In addition, we observe that with increasing number of RIS elements, the average transmit power is reduced for all the considered schemes utilizing RIS. Furthermore, we can observe a performance gain of the proposed solution of around $4$~dB and $6$~dB at $N=64$ compared to conventional SLP (SLP infinite, with RIS) and the SLP finite, no RIS scheme, respectively. This validates the benefits of RIS deployment in the proposed design.

\subsection{Analysis of average transmit power as a function of the number of users}
\begin{figure}[h!]
	\centering	\includegraphics[width=0.45\textwidth]{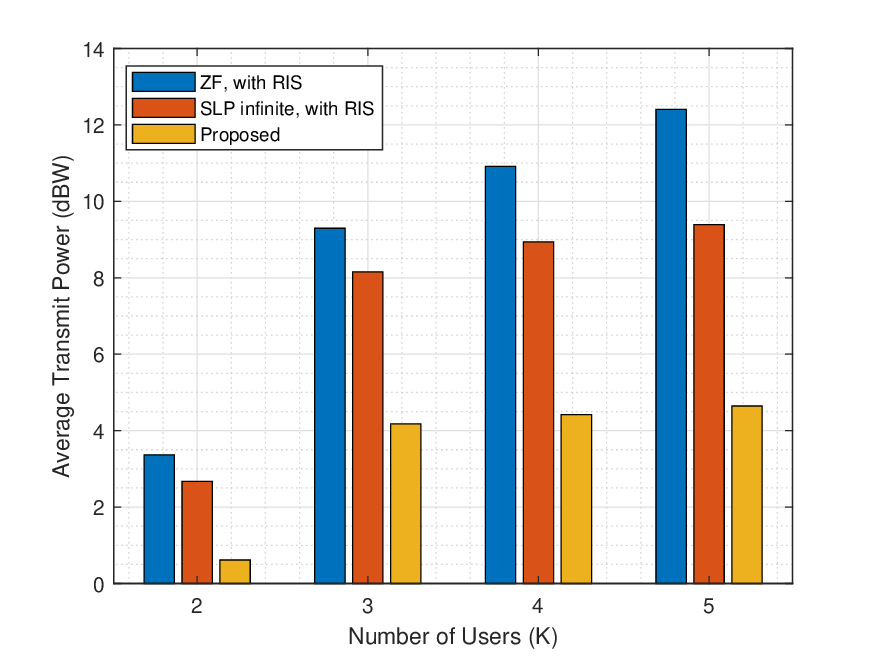}
	\caption{Average transmit power as a function of the number of users}
	\label{fig:C}
\end{figure}
This subsection studies the effect of the number of users on the average total transmit power. We vary the number of users from $2$ to $5$ and the block length is set at $L=8$. With increasing number of users, we note that the total transmit power increases. However, it is important to note that the performance gap between the proposed scheme and the benchmarks (ZF with RIS and SLP infinite with RIS) also increases as the number of users increases. This is mainly because the increased MUI  can be effectively exploited via the joint optimization of SLP, RIS phase shifts and constellation rotation. At $K=2$, we observe power savings of $2$~dB by the proposed solution compared to  conventional SLP (SLP infinite, with RIS) and $2.75$~dB with ZF precoding (ZF with RIS) scheme. This performance gap increases to $4.5$~dB and $6.4$~dB with conventional SLP and ZF precoding, respectively, at $K=4$. Further, we observe a performance gain of $4.9$~dB and $7.9$~dB of the proposed against the conventional SLP and ZF precoding, respectively, at $K=5$. The proposed solution becomes more and more effective as the number of users increases since there is a higher chance of exploiting constructive interference. Specifically, with increasing number of users the total number of symbol combinations that can occur in a packet increases exponentially. Accordingly, the choice of the constellation rotation becomes more and more important for the system performance. This is due to the broken symmetry of the symbol constellation as not all symbol constellations are equally probable as explained in \cite{kisseleff2021symbol}. Thus, the proposed scheme holds potential to facilitate the development of future large multiuser networks, while also providing significant energy efficiency gains.

\section{Conclusion}\label{sec:Conclusion}
In this paper, we studied the problem of total transmit power minimization in RIS-assisted SLP in the finite block length regime with constellation rotation. Specifically, total transmit power was minimized via the optimization of SLP, RIS phase shift design, and constellation rotation. Due to the non-convexity of the formulated problem, we resorted to alternating optimization to solve the joint optimization of SLP and RIS phase shift. Next, the optimal phase of the constellation rotation was obtained via an exhaustive search. Numerical results demonstrated the effectiveness of the proposed design in minimizing power consumption as compared to conventional SLP without constellation rotation, the design without RIS and zero forcing. We observed that with shorter block lengths, the proposed design is highly beneficial in obtaining significant power savings compared to the conventional RIS-assisted precoding design.

To further enhance the proposed design, future work will resort to machine learning to address the complexity of the proposed solution. Further, the case of imperfect CSI motivates the use of robust optimization,  which would make the design applicable to scenarios with non-negligible user mobility.
\bibliographystyle{IEEEtran}
\bibliography{main}
\end{document}